\documentclass[english, prl, amsmath, amssymb, superscriptaddress, twocolumn] {revtex4-2}

\usepackage{graphicx}
\usepackage{dcolumn}
\usepackage{bm}
\usepackage{lipsum}
\usepackage{hyperref}
\usepackage{layouts}
\usepackage{braket}
\usepackage[dvipsnames]{xcolor}
\usepackage[normalem]{ulem}
\usepackage{enumitem}
\usepackage{babel}
\usepackage{natbib}
\usepackage[separate-uncertainty=true, multi-part-units=single]{siunitx}
\usepackage{xcolor}

\DeclareSIUnit{\belmilliwatt}{Bm}
\DeclareSIUnit{\dBm}{\deci\belmilliwatt}

\makeatletter
\newcommand\thefontsize[1]{{#1 The current font size is: \f@size pt\par}}
\newcommand{\suppinf}[1]{(see Supplementary Information #1)}
\newcommand{\apdx}[1]{(see Appendix #1)}

\newcommand{\blender}{a}
\newcommand{\sem}{a inset}
\newcommand{\stabilityIctTwo}{b}
\newcommand{\magnetoSpertroscopyIctTwo}{c}
\newcommand{\stabilityIctThree}{d}
\newcommand{\magnetoSpertroscopyIctThree}{e}

\newcommand{\spectroIctTwoSingleExp}{a}
\newcommand{\spectroIctTwoSingleTh}{b}
\newcommand{\spectroIctTwoMultiExp}{c}
\newcommand{\zeroFieldEnergyDiagramIctTwo}{d}

\newcommand{\spectroIctOneExp}{a}
\newcommand{\spectroIctOneTh}{b}
\newcommand{\finiteFieldEnergyDiagramIctOne}{c}

\newcommand{\spectroIctThree}{a}
\newcommand{\finiteFieldEnergyDiagramIctThree}{b}

\newcommand{\vgOne}{V_{\rm{G1}}}
\newcommand{\vgTwo}{V_{\rm{G2}}}
\newcommand{\OneOne}{(1,1)}
\newcommand{\TwoZero}{(2,0)}
\newcommand{\OneZero}{(1,0)}
\newcommand{\ZeroOne}{(0,1)}
\newcommand{\singletOneOne}{S(1,1)}
\newcommand{\singletTwoZero}{S(2,0)}

\newcommand{\tripletZero}{T$_0$(1,1)}

\newcommand{\upUp}{$\ket{\Uparrow \Uparrow}$}
\newcommand{\upDown}{$\ket{\Uparrow \Downarrow}$}
\newcommand{\downUp}{$\ket{\Downarrow \Uparrow}$}
\newcommand{\downDown}{$\ket{\Downarrow \Downarrow}$}
\newcommand{\leftdot}{$\ket{\mathrm{L}}$}
\newcommand{\leftDown}{$\ket{\mathrm{L} \Downarrow}$}
\newcommand{\leftUp}{$\ket{\mathrm{L} \Uparrow}$}
\newcommand{\rightdot}{$\ket{\mathrm{R}}$}
\newcommand{\rightDown}{$\ket{\mathrm{R} \Downarrow}$}
\newcommand{\rightUp}{$\ket{\mathrm{R} \Uparrow}$}

\newcommand{\deltaSO}{\Delta_{\rm{SO}}}
\newcommand{\gc}{g_{\rm{c}}}
\newcommand{\wRabi}{\omega_{\rm{R}}}

\newcommand{\wRabiZero}{\omega_{\rm{R0}}}
\newcommand{\fRes}{f_{\rm{r}}}
\newcommand{\wRes}{\omega_{\rm{r}}}
\newcommand{\fExc}{f_{\rm{exc}}}
\newcommand{\epsilonSO}{\varepsilon_{\rm{SO}}}

\newcommand{\fResonator}{497}

\newcommand{\ictOneDeltaSO}{4.6 \pm 0.1}
\newcommand{\ictOnegTwo}{1.92 \pm 0.02}
\newcommand{\ictOnegOneLimit}{2.38}

\newcommand{\ictTwoAlphaTwo}{0.160 \pm 0.001}
\newcommand{\ictTwoAlphaOne}{0.50 \pm 0.02}
\newcommand{\ictTwoDelta}{5.72 \pm 0.04}

\newcommand{\ictThreeDeltaSO}{2.4}

\newcommand{\ictThreeGright}{1.33}
\newcommand{\ictThreeGleft }{1.27}

\DeclareMathOperator{\resp}{\chi}

\makeatother

\begin{document}

\title{Dispersively probed microwave spectroscopy of a silicon hole double quantum dot}

\author{Rami Ezzouch}
\email{rami.ezzouch@cea.fr}
\affiliation{Univ. Grenoble Alpes, Grenoble INP, CEA, IRIG-PHELIQS, F-38000 Grenoble, France}
\author{Simon Zihlmann}
\affiliation{Univ. Grenoble Alpes, Grenoble INP, CEA, IRIG-PHELIQS, F-38000 Grenoble, France}
\author{Vincent P. Michal}
\affiliation{Univ. Grenoble Alpes, CEA, IRIG-MEM, F-38000 Grenoble, France}
\author{Jing Li}
\affiliation{Univ. Grenoble Alpes, CEA, IRIG-MEM, F-38000 Grenoble, France}
\author{Agostino Apr\'a}
\affiliation{Univ. Grenoble Alpes, Grenoble INP, CEA, IRIG-PHELIQS, F-38000 Grenoble, France}
\author{Benoit Bertrand}
\affiliation{CEA, LETI, Minatec Campus, F-38000 Grenoble, France}
\author{Louis Hutin}
\affiliation{CEA, LETI, Minatec Campus, F-38000 Grenoble, France}
\author{Maud Vinet}
\affiliation{CEA, LETI, Minatec Campus, F-38000 Grenoble, France}
\author{Matias Urdampilleta}
\affiliation{Univ. Grenoble Alpes, CNRS, Grenoble INP, Institut N\'eel, F-38000 Grenoble, France}
\author{Tristan Meunier}
\affiliation{Univ. Grenoble Alpes, CNRS, Grenoble INP, Institut N\'eel, F-38000 Grenoble, France}
\author{Xavier Jehl}
\affiliation{Univ. Grenoble Alpes, Grenoble INP, CEA, IRIG-PHELIQS, F-38000 Grenoble, France}\author{Yann-Michel Niquet}
\affiliation{Univ. Grenoble Alpes, CEA, IRIG-MEM, F-38000 Grenoble, France}
\author{Marc Sanquer}
\affiliation{Univ. Grenoble Alpes, Grenoble INP, CEA, IRIG-PHELIQS, F-38000 Grenoble, France}\author{Silvano De Franceschi}
\email{silvano.defranceschi@cea.fr}
\affiliation{Univ. Grenoble Alpes, Grenoble INP, CEA, IRIG-PHELIQS, F-38000 Grenoble, France}\author{Romain Maurand}
\email{romain.maurand@cea.fr}
\affiliation{Univ. Grenoble Alpes, Grenoble INP, CEA, IRIG-PHELIQS, F-38000 Grenoble, France}

\date{\today}

\begin{abstract}
\textbf{Owing to ever increasing gate fidelities and to a potential transferability to industrial CMOS technology, silicon spin qubits have become a compelling option in the strive for quantum computation. In a scalable architecture, each spin qubit will have to be finely tuned and its operating conditions accurately determined. In this prospect, spectroscopic tools compatible with a scalable device layout are of primary importance. Here we report a two-tone spectroscopy technique providing access to the spin-dependent energy-level spectrum of a hole double quantum dot defined in a split-gate silicon device. A first GHz-frequency tone drives electric-dipole spin resonance enabled by the valence-band spin-orbit coupling. A second lower-frequency tone ($\approx \SI{500}{\mega\hertz}$) allows for dispersive readout via rf-gate reflectometry. We compare the measured dispersive response to the linear response calculated in an extended Jaynes-Cummings model and we obtain characteristic parameters such as g-factors and tunnel/spin-orbit couplings for both even and odd occupation.
}
\end{abstract}

\maketitle




\section{\label{sec:intro}Introduction}

Single spins localized in gate defined quantum dots have been recognized as a promising platform for quantum computation early on \cite{Loss1998}. Among them, spin qubits in silicon have attracted a lot of interest due to their long coherence \cite{Yoneda2018a} and possibility of single \cite{Veldhorst2014} and two-qubit gates \cite{Veldhorst2015}. The steady evolution towards increasing the number of qubits has triggered the quest for a compact control and read-out architecture. Considering qubit control, all electrical qubit driving appears as a clear asset in dense quantum dot array and has been investigated with intrinsic or artificial spin-orbit interaction either in silicon \cite{Kawakami2014, Yoneda2018a, Maurand2016} or germanium \cite{Watzinger2018,Hendrickx2020c,Hendrickx2020a}. For readout, the commonly used spin-to-charge conversion with subsequent charge detection through nearby charge sensors is becoming increasingly challenging as the number of quantum dots to sense increases \cite{Mortemousque2018, Hendrickx2020b}. Gate-based dispersive readout \cite{Cottet2011, Colless2013} is an alternative strategy possibly usable for a dense array \cite{Zheng2019}. It benefits in particular from large gate lever arms \cite{Gonzalez-Zalba2015, Ahmed2018, Crippa2019} and single shot read-out has been demonstrated \cite{Pakkiam2018, Urdampilleta2019, West2019} with fidelity up to \SI{98}{\percent} for \SI{6}{\micro\second} integration time with on-chip resonators \cite{Zheng2019}.
Despite all the advantages of gate-based dispersive readout, it also has its limitations. One of the main drawbacks is the limited information about the fundamental energy spectrum of each qubit gate-based dispersive readout is able to provide compared to commonly used charge sensing. This includes trivial quantities such as gate lever arms and charge tunnel couplings in general but also quantum dot level spacings and in particular for spin-orbit qubits the g-factor and spin-orbit interaction strength. However, dispersive readout coupled with microwave spectroscopy has been proven to be a powerful tool to access some of these information, among them lever arm \cite{Penfold-Fitch2017} and charge tunnel coupling \cite{Petersson2010, Penfold-Fitch2017}.

With the rapid progress of hole spin qubits in silicon \cite{Maurand2016, Crippa2018, Crippa2019} and germanium \cite{Watzinger2018, Hendrickx2020a, Hendrickx2020b, Hendrickx2020c, Froning2020}, there is a strong need to characterize and explore these qubits. Not only do hole spin qubits allow for a dense integration of qubits \cite{Hendrickx2020b} while preserving individual addressability \cite{Maurand2016, Lawrie2020}, but there is plenty of physics in the interplay of orbital and spin degree of freedom to discover. 

Here we perform a two-tone magneto-spectroscopy on a hole silicon double quantum dot (DQD) as a model system. Due to spin-orbit interaction, spin-flip photon assisted tunneling is possible under microwave irradiation and detectable by gate-based dispersive readout. The spectroscopy allows then the reconstruction of the entire energy spectrum of the DQD necessary for qubit control and readout. A qubit driven Jaynes-Cummings Hamiltonian resolved in the linear regime is supporting our experimental findings.

\section{\label{sec:methods}Methods}

\begin{figure}[!htbp]
\centering
\includegraphics[width=\columnwidth]{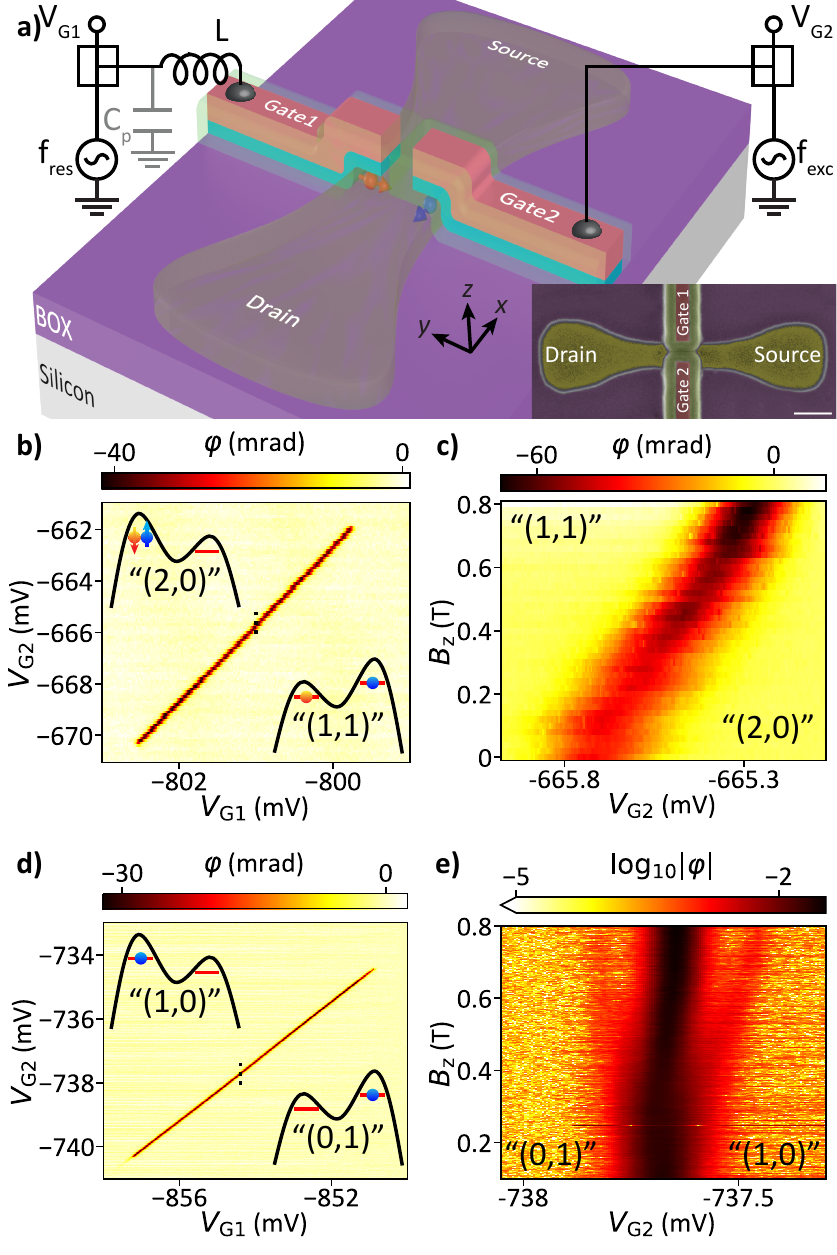}
    \caption{\label{fig:fig1} \textbf{Double quantum dot device \& working points.} \textbf{(a)} Simplified 3D schematic of a split-gate, silicon-on-insulator field-effect transistor. A LC resonator wired to gate~1 is used for reflectometry readout. Static voltages $\vgOne{}$ and $\vgTwo{}$ are applied to gates 1 and 2. Using bias tees, they are combined with a $\sim$\SI{500}{\mega\hertz} reflectometry tone ($\fRes{}$) and a 1-\SI{20}{\giga\hertz} microwave spectroscopy tone ($\fExc{}$), respectively. The inset shows a false color scanning electron micrograph of the device (scale bar: \SI{100}{\nano\meter}). \textbf{(b), (d)} Phase response of the LC resonator as a function of $\vgOne{}$ and $\vgTwo{}$ showing inter-dot charge transitions ICT~2 and ICT~3 for even and  odd parity, respectively. The insets show the equivalent one- and two-electron charge configurations just above and below ICT~3 and ICT~2, respectively. The first (second) number represents the equivalent hole occupation in the dot under gate~1 (gate~2). \textbf{(c), (e)} Phase response as a function of $\vgTwo{}$ and $B_{\rm{z}}$ at fixed $\vgOne{}$ (see dashed line in (b) and (d)) revealing the ground-state evolution in magnetic field.}
\end{figure}

The experiment is realized in a p-type double gate MOSFET fabricated on a \SI{300}{\milli\meter} silicon-on-insulator wafer using an industry-standard fabrication line \cite{Voisin2014, Maurand2016}. The silicon channel is nominally \SI{10}{\nano\meter} thick, \SI{80}{\nano\meter} wide, and is partially overlapped by two \SI{32}{\nano\meter} long gates (see Fig.~\ref{fig:fig1}\sem{} for an electron micrograph of a nominally identical device). The gates are defined by e-beam lithography and have enlarged SiN spacers to avoid doping implantation in the channel. This face-to-face geometry allows the accumulation of a hole DQD in parallel (with source and drain) by applying negative DC voltages $\vgOne{}$ and $\vgTwo{}$ on the gates when the device is operated in a dilution refrigerator at the base temperature $T_{\rm{base}}=\SI{30}{mK}$. The two dots are formed at the corners of the channel overlapped by the gates \cite{Voisin2014}. Fig.~\ref{fig:fig1}\blender{} schematically shows the device as well as its gate connections. Gate~2 as well as the drain (not shown) are connected to a broadband high frequency coaxial line that allows to perform microwave two tone spectroscopy. Gate~1 is connected to a surface mount inductor ($L=\SI{220}{\nano\henry}$), which forms with its parasitic capacitance and device impedance an LC resonator with a resonance frequency $\fRes{}=\wRes{}/(2\pi)=\SI{\fResonator{}}{\mega\hertz}$, a loaded quality factor $Q_{\rm{loaded}} \approx 24$ and a characteristic impedance $Z_c\approx\SI{600}{\ohm}$ (see Supplementary Information A 
and B
for a complete measurement setup description and resonator characterization).\\

The hole DQD acts as a variable load for the $LC$ resonator and the resonance frequency undergoes a dispersive shift depending on the DQD state. This can be readily understood if one considers a (M+1, N) $\leftrightarrow$ (M, N+1) interdot charge transition with M (N) the charge number in the left (right) dot. The two-state Hamiltonian of this system writes $H_{\rm{DQD}} = -\varepsilon\,\sigma_{\rm{z}}/2 - \Delta\,\sigma_{\rm{x}}/2$, where the Pauli matrices act in the space of the charge configuration. $\sigma_{\rm{x}}$ describes the tunneling between the dots that opens a gap $\Delta$ in the energy spectrum. The difference in energy of the two states reads $E = \sqrt{\varepsilon^2 + \Delta^2}$ and is a function of the detuning $\varepsilon$. In the adiabatic limit, when the resonator angular frequency $\wRes \ll E/\hbar$, the interaction between a charge qubit and a resonator has often been treated semi-classically with the introduction of quantum capacitances \cite{Duty2005, Sillanpaa2005, Mizuta2017}. However, when the frequency of the readout oscillator is comparable to the characteristic frequency of the measured system ($\wRes \simeq E/\hbar$), a quantum mechanical treatment of the interaction with an extended Jaynes-Cummings Hamiltonian is convenient in order to take into account the finite frequency of the readout apparatus. 
Recently, it was shown that for a charge qubit, such a model captures the interaction also in the adiabatic limit \cite{Park2020}. Here, we therefore model all the interactions within the framework of the Jaynes-Cummings Hamiltonian. This quantum approach provides complementary physical insights and proves useful in describing more complex situations. In particular, we extend the Jaynes-Cummings model to the driven case (with microwaves at frequency $\fExc{}$), where we also capture dispersive shifts due to resonantly driven transitions between states in the DQD (a complete discussion of the underlying theory can be found in Appendix \ref{apdx:theory}).

The coupling between the DQD and the read-out resonator, described by the Hamiltonian $H_{\rm{r}}=\hbar\wRes\left(a^\dagger a+1/2\right)$, is expressed in the basis of the charge configurations as $H_{\rm{int}} = \hbar \gc \sigma_{\rm{z}}\left(a+a^\dagger\right)$, with $a$ the annihilation operator of the oscillator, $\gc=(\alpha_1-\alpha_2)eV_{\rm{rms}}/(2\hbar)$ the coupling strength between the charge and the microwave photons, $V_{\rm{rms}}=\sqrt{\hbar\wRes/(2C_{\rm{r}})}$ the zero-point voltage fluctuation of the $LC$ oscillator, and $C_{\rm{r}}$ is the capacitance of the $LC$ circuit (which includes the geometric capacitance of the DQD). The coupling between gate~1 and the double quantum dot charge leads to a phase shift between the incoming and the reflected microwaves. In the linear regime discussed in Appendix \ref{apdx:theory}, the phase shift can be expressed as
\begin{equation}
    \label{eq:phase}
    \delta\phi=\frac{4Q_{\rm loaded}{\rm Re}\chi(\wRes)}{\wRes},
\end{equation}
where $\chi(\wRes)$ is the charge response function whose real part represents the linear shift in the resonant angular frequency of the $LC$ circuit. For a pair of states near charge degeneracy with the readout oscillator in the adiabatic limit, the response function is real and equals
\begin{equation}
    \label{eq:dispersive_shift_two_state}
    \chi = -\frac{2\hbar \gc^2 \Delta^2}{\left(\varepsilon^2 + \Delta^2\right)^{3/2}};\,k_B T,\,\hbar\wRes,\,\hbar\Gamma_2 \ll \Delta,
\end{equation}
where $k_B$ is the Boltzmann constant, $T$ is the equilibrium temperature of the DQD environment and $\Gamma_2$ is the decoherence rate of the DQD charge. Eq.~\ref{eq:dispersive_shift_two_state} is equivalent to the standard oscillator shift $\delta \wRes=-C_{\rm{Q}}\wRes/(2C_{\rm{r}})$, with $C_{\rm{Q}}$ the quantum capacitance of the DQD \cite{Petersson2010, Colless2013, Mizuta2017}.

\section{\label{sec:results}Results}
When measuring the phase response of the resonator at its resonance frequency while sweeping the gate voltages $\vgOne{}$ and $\vgTwo{}$, we obtain the charge stability diagram of the DQD system \suppinf{C}. 
Diagonal features with positive slope in this diagram mark interdot charge transitions (ICTs). This work is focused on three interdot transitions ICT~1 and ICT~2 on device~1 and ICT~3 on device~2, all chosen to have an estimated hole number below 20 in each dot. Fig.~\ref{fig:fig1}\stabilityIctTwo{} and \ref{fig:fig1}\stabilityIctThree{} show the stability diagrams around ICT~2 and ICT~3 respectively. The same figures for the case of ICT~1 are included in the Supplementary Information C. 
Using the model introduced above and fitting the phase response as a function of detuning at each ICT we find a charge-photon coupling $\gc/(2\pi)\simeq\SI{35}{\MHz}$ for all three interdot transitions, see Appendix \ref{apdx:gcoupling}.

Any given ICT is characterized by either an even or an odd parity of the total occupation number in the DQD. Without knowing this number, different parities can still be discriminated through the magnetic field evolution of the corresponding ICT phase response \cite{Cottet2011, Schroer2012, Urdampilleta2015}.


\subsection{\label{subsec:results:interdot}Parity of the DQD}
For holes in silicon, the presence of spin-orbit interaction is changing the magnetic field dependence of the ICT as spin-flip tunneling is allowed and couples different spin states. However, we show in the next paragraph that it is still possible to infer the interdot charge parity from the dispersive response in magnetic field. The phase response of the LC resonator is measured as a function of $\vgTwo{}$ and magnetic field, keeping $\vgOne{}$ constant. 
Measurements are shown in Fig.~\ref{fig:fig1}\magnetoSpertroscopyIctTwo{} and \ref{fig:fig1}\magnetoSpertroscopyIctThree{}.

In the even case, Fig.~\ref{fig:fig1}\magnetoSpertroscopyIctTwo{}, the dip in phase at zero detuning and zero magnetic field arising from the avoided crossing of the \singletOneOne{}~$\leftrightarrow$~\singletTwoZero{} remains unchanged as long as the Zeeman energy of the polarized triplet state $E_{\rm{Z}}<\Delta$. Once $E_{\rm{Z}}$ becomes larger than $\Delta$, the state \downDown{} becomes the ground state in the \OneOne{} configuration and a new avoided-crossing mediated by spin-orbit interaction between \downDown{} and \singletTwoZero{} emerges, which we characterize by an energy gap $\deltaSO{}$, see Fig.~\ref{fig:fig3}c for an energy diagram at finite magnetic field. With increasing magnetic field, this avoided crossing moves towards higher detuning, which explains why the dip in phase in Fig.~\ref{fig:fig1}\magnetoSpertroscopyIctTwo{} moves towards larger $\vgTwo{}$ as the magnetic field is increased. Moreover, the increase in phase shift is due to $\deltaSO{} < \Delta$, which gives rise to a higher dispersive shift $\chi$ at higher field following Eq.~\ref{eq:dispersive_shift_two_state}.

In the odd case, see Fig.~\ref{fig:fig1}\magnetoSpertroscopyIctThree{}, the central dip in phase does not vary much with increasing magnetic field indicating that the nature of the ground state is unchanged. However Fig.~\ref{fig:fig1}\magnetoSpertroscopyIctThree{} shows two additional phase signals appearing on either side of the central phase dip. These originate from higher-lying avoided crossings in the DQD level spectrum that lead to nonzero electric susceptibility, see Fig.~\ref{fig:fig4}\finiteFieldEnergyDiagramIctThree{} for an energy diagram at finite magnetic field. Their origin is the spin-orbit mediated coupling of states with opposite spins in the two dots. With increasing magnetic field, these dips in phase move away from the central feature and fade out. The dispersion is again linked to a change in $E_{\rm{Z}}$, whereas the reduced phase signal is explained by a lower occupation probability of the excited state by thermal activation. The slight dispersion of the central dip arises from a difference in the Land\'e g-factors of the two QDs. The dispersive detection of higher-lying avoided crossings due to spin-orbit interaction is similar to the recently reported observation of valley splittings in cavity-coupled electron quantum dots in silicon \cite{Mi2017}.

\subsection{\label{subsec:results:B0}Spectroscopy at zero magnetic field}

Having established the parity of each ICT, we proceed to microwave spectroscopy to explore the full DQD level spectrum as a function of magnetic field, $B_z$, along the $z$-axis, $i.e.$ perpendicular to the substrate. 
Both even and odd charge configurations have two anti-crossing states, with single and doublet spin character, respectively. 
The even configurations are characterized by the additional presence of spin-triplet states, which can be neglected at zero magnetic field because of their negligible dispersive shift. As a result, in both even and odd cases, a single charge tunnels between the two QDs giving rise to a nonzero electric susceptibility at the ICT, see Eq.~\ref{eq:dispersive_shift_two_state}. We can extend this model to capture microwave photon induced tunnel events by adding a fast electrical drive $A_{\rm{exc}}\cos\left(\omega_{\rm{exc}}t\right)/2$ to gate~2. The Hamiltonian describing the full system includes now also $H_{\rm{exc}} = A_{\rm{exc}}\cos\left(\omega_{\rm{exc}}t\right)\sigma_{\rm{z}}/2$, where $A_{\rm{exc}}$ is the amplitude of the fast drive signal. By solving the complete Hamiltonian $H = H_{\rm{DQD}}+H_{\rm{exc}} + H_{\rm{r}} + H_{\rm{int}}$, we find a linear response function
\begin{equation}
    \label{eq:dispersive_shift_two_state_driven}
    \begin{aligned}
    \chi\left(\wRes\right) ={} & -\left[-\frac{2\hbar \gc^2\Delta^2}{E^3}\frac{\delta\omega}{\wRabi} \right. \\
    & + \left. \left(\gc\frac{\varepsilon}{E}\frac{\wRabiZero}{\wRabi}\right)^2\frac{1}{\wRabi - \wRes - i\tilde{\Gamma}_2}\right]D,
    \end{aligned}
\end{equation}
where $\delta\omega = \omega_{\rm{exc}} - E/\hbar$ is the detuning of the drive frequency $\omega_d$ from the DQD transition energy $E$, $\wRabi = \sqrt{\delta\omega^2 + \wRabiZero^2}$ is the Rabi frequency due to the resonant drive with $\wRabiZero = A_{\rm{exc}}\Delta/(2\hbar E)$, $\tilde{\Gamma}_2$ is the decoherence rate of the driven system and $D$ is the difference of the occupation probability of the ground and excited state in the dressed basis. The first term is the adiabatic response already described in Eq.~\ref{eq:dispersive_shift_two_state}, whereas the second term includes now the dispersive shift due to resonantly driven charge transitions in the DQD. Let us note that in Eq.~\ref{eq:dispersive_shift_two_state_driven} we have only retained the rotating wave approximation (RWA) contribution of the resonant term which is relevant to our regime of operations. The region of accuracy of the RWA is quite extended, over a dot detuning range at least the gap $\Delta$, and matches the important region of the resonant signal. In other conditions where $\omega_r\ll\omega_R$ the non-RWA contribution may also be included straightforwardly. A detailed derivation of Eq.~\ref{eq:dispersive_shift_two_state_driven} is given in Appendix~\ref{apdx:theory}.


Figure ~\ref{fig:fig2}\spectroIctTwoSingleExp{} shows the phase response of the even-parity ICT~2 as a function of $\vgTwo{}$ and $\fExc{} = \omega_{\rm{exc}}/(2\pi)$ for fixed $\vgOne{}$. Qualitatively similar results can be found for odd parity cases, such as ICT~3 (not shown). To analyze the data in Fig. ~\ref{fig:fig2}\spectroIctTwoSingleExp{} we refer to the corresponding energy diagram at zero magnetic field, shown in Fig.~\ref{fig:fig2}\zeroFieldEnergyDiagramIctTwo{}. Resonant microwave induced transitions are highlighted by double arrows at positive and negative detuning.

\begin{figure}[!htbp]
\centering
\includegraphics[width=\columnwidth]{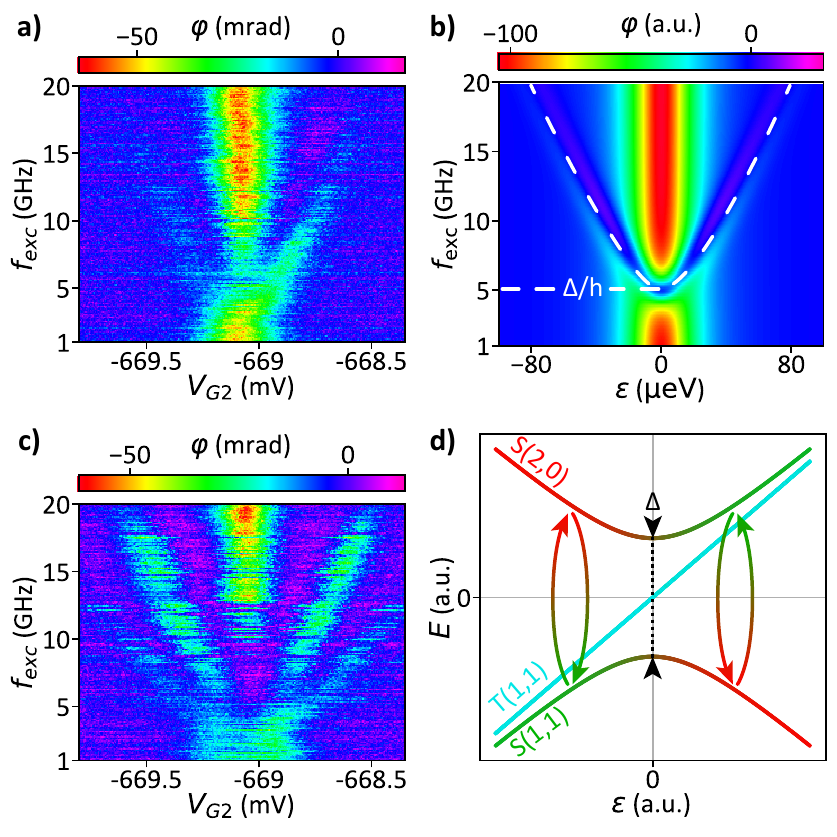}
    \caption{\label{fig:fig2}\textbf{Photon assisted spectroscopy at zero magnetic field.} \textbf{(a)} Phase response of the resonator as a function of gate voltage, $\vgTwo{}$, and microwave frequency, $\fExc$, at zero  magnetic field. The output power of the microwave generator is adjusted for each $\fExc$ in order to deliver a constant power at the device \suppinf{D}. 
    In addition to the central interdot transition signal vanishing at \SI{\ictTwoDelta{}}{\giga\hertz}, two side branches mark photon-assisted charge transitions between the quantum dots. \textbf{(b)} Theoretical simulation of the driven DQD phase response. The central dip at $\varepsilon = 0$ vanishes when the excitation energy matches $\Delta$. \textbf{(c)} Multi-photon processes arising at higher driving power. Additional side branches appear at one half and one third of the side-branch frequency in (a) indicating two-photon and three-photon process, respectively. \textbf{(d)} Energy diagram of a DQD near the ``\OneOne{}''~$\leftrightarrow$~``\TwoZero{}'' transition at zero magnetic field. The double arrows mark the processes giving rise to the branches observed in (a).
    }
\end{figure}

The vertical ridge at $\vgTwo{} \simeq \SI{-669.1}{\milli\volt}$\textbf{} corresponds to the dispersive shift arising from the charge qubit associated to the \singletOneOne{}-\singletTwoZero{} anticrossing as described by the first term in Eq.~\ref{eq:dispersive_shift_two_state_driven}. The phase dip along this line vanishes when the microwave excitation energy matches the energy gap $\Delta$ due to tunnel coupling, $i.e.$ when $\delta\omega \rightarrow 0$ \cite{Petersson2010,Penfold-Fitch2017, Urdampilleta2015}. From this we find $\Delta/h =$~\SI{\ictTwoDelta{}}{\giga\hertz} \apdx{\ref{apdx:tunnel}}

Two side branches can be seen in the the same figure. They consist of dip-peak features due to microwave-assisted excitation away from the charge degeneracy point at $\epsilon = 0$. They occur when the microwave tone is in resonance with the charge qubit energy $E$, once again when $\delta\omega\rightarrow 0$. and they are accounted for by the second term in Eq.~\ref{eq:dispersive_shift_two_state_driven}.


At large detuning ($\varepsilon\gg\Delta$), the side branch turn into straight lines whose slope can be used to extract the lever-arm parameter $\alpha_2=\ictTwoAlphaTwo{}$ relating detuning energy to gate~2 voltage, see Appendix \ref{apdx:alpha}. From the slope of the ICT~2 line in Fig.~\ref{fig:fig1}\stabilityIctTwo{}, we can further infer the lever-arm parameter for gate~1, $\alpha_1=\ictTwoAlphaOne{}$.

The data in Fig.~\ref{fig:fig2}\spectroIctTwoSingleExp{} shows a clear asymmetry between the two branches, with the branch at positive detuning being more pronounced. 
This asymmetry may be ascribed to the presence of the triplet states, which affects the population of the anti-crossing singlet states. At negative detuning, the \singletOneOne{} ground state is partially depopulated due the thermal population of the closely lying triplet excited states. This should lead to fainter side branch since the triplets do not contribute any measurable dispersive phase shift and cannot be photon-excited to the \singletTwoZero{} state due to time-reversal symmetry at zero magnetic field. 


Using the model introduced above, we can qualitatively reproduce the experimental results, as shown in Fig.~\ref{fig:fig2}\spectroIctTwoSingleTh{}, where the transition energy, $E$, is highlighted by a white dashed line. The exact shape of the dip-/peak phase branches that emerge when the charge qubit is driven resonantly at non-zero detuning is a sensitive function of dephasing and relaxation whose complete modelling goes beyond the scope of this work.

When the system is strongly driven , multi-photon processes can occur ($n h \fExc{} = E$, where $n$ is an integer). Fig.~\ref{fig:fig2}\spectroIctTwoMultiExp{} shows the phase response at large microwave power. In addition to the dispersive shift of the driven charge qubit originating from a one photon process, new branches appear at half and one-third the frequencies of the original branches demonstrating two- and three-photon processes, respectively. The theoretical description of the multi-photon case can be found in Appendix \ref{apdx:multiphoton}.

To sum up, microwave spectroscopy at zero magnetic field is a powerful tool to extract the interdot charge tunnel coupling as well as the lever-arm parameters for both gates allowing the reconstruction of the DQD spectrum at zero magnetic field.

\subsection{\label{subsec:results:Bfinite} Spectroscopy of two holes in a DQD at finite magnetic field}
\begin{figure}[!htbp]
\centering
\includegraphics[width=\columnwidth]{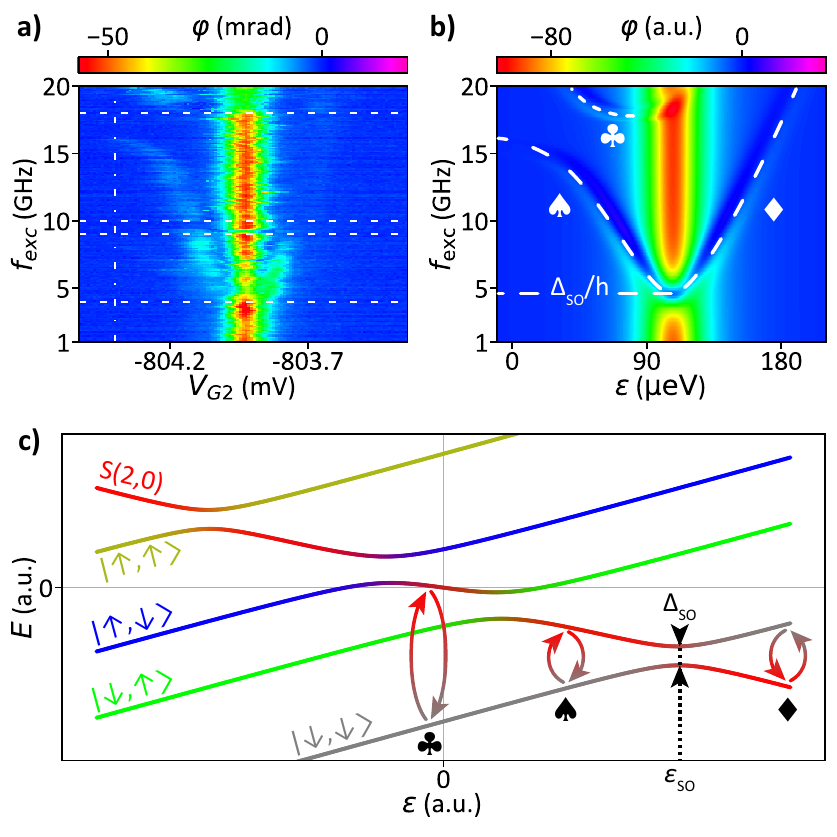}
    \caption{\label{fig:fig3}\textbf{Photon assisted spectroscopy at finite magnetic field for an even-parity inter-dot charge transition.} \textbf{(a)} Phase response of the LC resonator around the spin-orbit anticrossing of \downDown{} with \singletTwoZero{} as a function of $\vgTwo{}$ and $\fExc$ at $B_{\rm{z}}=\SI{600}{\milli\tesla}$. The dashed horizontal lines delimit regions in which the spectroscopy tone power is held constant at room temperature. The dash-dotted vertical line marks position of the ICT at zero magnetic field, corresponding to  $\varepsilon = 0$. The strong vertical structure is associated with the spin-orbit anticrossing at $\varepsilon = \epsilonSO{}$. It vanishes at 
    $\fExc = \deltaSO{}/h = $\SI{\ictOneDeltaSO{}}{\giga\hertz}. The three side branches around the central signal correspond to photon induced charge transitions between the quantum dots, as indicated in (c). \textbf{(b)} Corresponding simulated phase response of the driven DQD.   \textbf{(c)} Energy diagram of DQD of around a ``\OneOne{}''~$\leftrightarrow$~``\TwoZero{}'' transition at finite magnetic field and with $g_1 \neq g_2$. The photon-induced charge transitions responsible for the side branches in (a) and (b) are indicated by arrows and corresponding symbols.
    }
\end{figure}

We now proceed with microwave spectroscopy at finite magnetic field to explore the spin-split energy levels and the spin-orbit coupling in the DQDs. First, we present results for ICT~2. At an external magnetic field $B_{\rm{z}}=\SI{600}{\milli\tesla}$, the triplet states split leading to a DQD energy spectrum as illustrated in Fig.~\ref{fig:fig3}\finiteFieldEnergyDiagramIctOne{}, where the \downDown{} is the ground state at $\varepsilon=0$. Due to a difference in $g$-factors between the two quantum dots, the \tripletZero{} state mixes with the singlet \singletTwoZero{} state around $\varepsilon = 0$ and a new basis for the \OneOne{} states consisting of four non-degenerate states \downDown{}, \upDown{}, \downUp{} and \upUp{} needs to be adopted. At positive detuning, \downDown{} and \singletTwoZero{} couple due to the intrinsic spin-orbit coupling in the valence band of silicon. At finite magnetic field, This gives rise to an avoided crossing $\deltaSO{}$ with characteristic energy $\deltaSO{}$ at a magnetic-field-dependent detuning $\varepsilon = \epsilonSO{}$.

Around $\epsilonSO{}$, the hole DQD could be operated as a ``spin-flip'' charge qubit. In Fig.~\ref{fig:fig3}\spectroIctOneExp{}, we show a two-tone spectroscopy around $\varepsilon = \epsilonSO{}$. The dispersive interaction of the qubit with the resonator gives rise to the vertical dip structure at $\vgTwo{} \simeq \SI{-803.9}{\milli\volt}$. As for the case of a spin-less charge qubit (Fig.~\ref{fig:fig2}), we observe a local suppression of this dip structure from which we extract a spin-orbit-mediated avoided crossing $\deltaSO{}/h = \SI{\ictOneDeltaSO{}}{\giga\hertz}$ \apdx{\ref{apdx:tunnel}}.

Away from $\epsilonSO{}$, the dispersive shift due to the driven ``spin-flip'' charge qubit arises when the microwave photon energy matches the energy splitting between \singletTwoZero{} and \downDown{}. 
Close to $\vgTwo{}=\SI{-804.2}{\milli\volt}$, $i.e.$ close to zero detuning, the left branch bends towards a horizontal asymptote around \SI{16}\,{GHz}. This arises from the hybridization of the exited state \singletTwoZero{} with the \downUp{} state, as shown in Fig.~\ref{fig:fig3}\finiteFieldEnergyDiagramIctOne{}. 
In this regime, the ``spin-flip'' charge qubit evolves to a single-dot ``spin-orbit'' qubit for which the electric dipole is largely reduced and cannot be sensed by the LC resonator.
From the frequency of the horizontal asymptote we extract the $g$-factor of the second dot, $g_2=\ictOnegTwo{}$. At $\vgTwo{}=\SI{-804.2}{\milli\volt}$ and for frequencies close to \SI{20}{\giga\hertz} an additional phase signal is visible in Fig.~\ref{fig:fig3}\spectroIctOneExp{}. This signal can be associated to the transition between \downDown{} and the hybridized \singletTwoZero{} and \upDown{} states. In principle, this branch could allow for the extraction of the $g$-factor of the first dot. However, due the upper limit of \SI{20}{\giga\hertz} in our microwave generator, we were not able to fully capture this feature and we only infer $g_1 > \ictOnegOneLimit{}$.

Similarly to the zero magnetic field case, the dispersive shift of the resonator can be modeled by also taking into account the spin degree of freedom as well as the spin orbit interaction. Apart from spin-flip tunnel events, the physics remains the same. We again find a qualitative agreement with the measurements, see Fig.~\ref{fig:fig3}\spectroIctOneTh{}. The white dashed lines highlight the transition energies as indicated with different arrows in Fig.~\ref{fig:fig3}\finiteFieldEnergyDiagramIctOne{}. We again note that the exact shape of the side wings depend on the details of the decoherence of the driven system and therefore exact modeling goes beyond the scope of this work.

\subsection{\label{subsec:results:Bfinite1hole}Spectroscopy of a single hole in a DQD at finite magnetic field}

\begin{figure}[!htbp]
\centering
\includegraphics[width=\columnwidth]{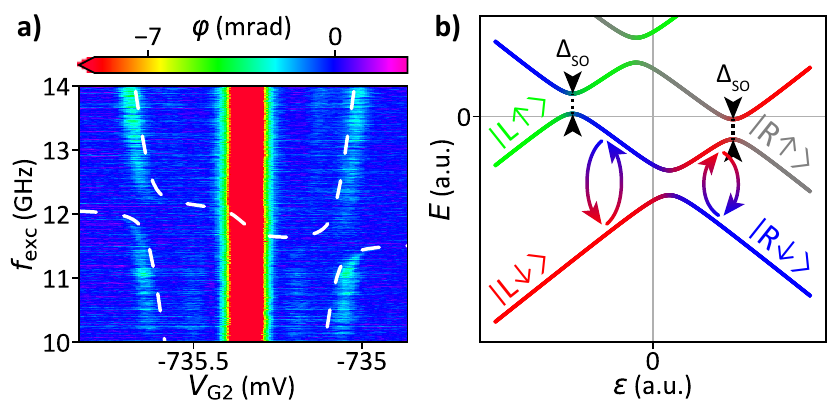}
    \caption{\label{fig:fig4}\textbf{Photon assisted spectroscopy at finite magnetic field for an odd-parity inter-dot charge transition.} \textbf{(a)} Phase response of the LC resonator around zero detuning as a function of $\vgTwo{}$ and $\fExc$ at $B_z =$~\SI{1.3}{\tesla}. The dispersive response due to the charge qubit is visible as the vertical feature at $\vgTwo{} \approx \SI{-737.3}{\milli\volt}$. At resonance, $2h\fExc{} = \Omega$, the resonator undergoes as well a phase shift (white dashed line). Around \SI{12}{\giga\hertz} spin-orbit anticrossings between states with opposite spin localized in different dots are detectable in the driven response. \textbf{(b)} Energy diagram of a single hole in a DQD (``\ZeroOne{}''~$\leftrightarrow$~``\OneZero{}'' transition) at finite magnetic field with $g_1 \simeq g_2$. At zero detuning, the down spin states of the left and right dot undergo an anticrossing due to tunneling $t$. A small anticrossing due to the spin-orbit interaction appears between the down spin states of one dot and the up spin states of the other dot. Microwave induced transitions that give rise to the two branches are highlighted with arrows.}
\end{figure}

We now present in Fig.~\ref{fig:fig4} microwave spectroscopy measurements for ICT~3, the odd-parity ICT. At finite magnetic field the basis states \leftdot{} and \rightdot{} of an odd parity ICT are spin split into \leftDown{}, \leftUp{}, \rightDown{} and \rightUp{}, resulting in the energy diagram of Fig.~\ref{fig:fig4}\finiteFieldEnergyDiagramIctThree{}. Around zero detuning, pure charge tunnel coupling gives rise to avoided crossings between states with the same spin, $i.e.$ between \leftDown{} and \rightDown{} and between \leftUp{} and \rightUp{}.
The strong central feature in Fig.~\ref{fig:fig4}\spectroIctThree{} is due to the dispersive shift associated with the lowest energy one, involving spin-down states. 

Similar to the even-parity case, two side branches arise when $h \fExc{}$ matches the energy difference between ground and excited states, corresponding to transitions from \leftDown{} to \rightDown{} and vice-versa. They exhibit clear avoided crossings around \SI{12}{\giga\hertz}. The one on the left (right) is due to a spin-orbit-mediated tunnel coupling between \leftUp{} and \rightDown{} (\leftDown{} and \rightUp{}). 
We would like to point out that the observed branches, highlighted by white dashed lines, are in fact due to two-photon excitations. As a consequence, transition frequencies 
in Fig.~\ref{fig:fig4}\spectroIctThree{} are a factor of two smaller than the actual transition energies.  
From the amplitude of the measured avoided crossings we find a spin-orbit gap 
$\deltaSO{}/h=\SI{\ictThreeDeltaSO{}}{\giga\hertz}$. 
In addition, the side branch on the left (right) approaches asymptotically a constant frequency set by the Zeeman energy in the left (right) quantum dot. This allows us to determine the two $g$-factors of the DQD, which happen to be differ slightly from each other, $i.e.$ $g_{\rm{L}} = \ictThreeGleft$ and $g_{\rm{R}} = \ictThreeGright$.

\section{\label{sec:conclusion}Conclusions}
In conclusion, we have performed microwave magneto-spectroscopy in combination with gate-based dispersive readout of silicon hole DQDs. By modelling the DQD coupled to the LC resonator and microwave spectroscopic tone by a driven Jaynes-Cummings Hamiltonian, we derive the linear response function of the system, which qualitatively explains our experimental data. Due to the spin-orbit interaction present in the valence band of silicon, all spin-orbit states have been revealed by two-tone spectroscopy enabling a precise reconstruction of the DQD energy diagram in both even- and odd-parity inter-dot transitions. Consequently, we were able to extract all of necessary physical parameters of a DQD have been extracted, $i.e.$ gate lever arms, tunnel couplings, g-factors and spin-orbit strength.
Combined the demonstrated two-tone spectroscopy with frequency multiplexed gate dispersive readout could enable parameter characterization in dense arrays of spin-orbit qubits without the need for local charge detectors and reservoirs whose integration is technically challenging. Moreover, the use of superconducting LC resonators  with higher resonance frequency, Q-factor and impedance, either off-chip \cite{Ibberson2020} or on-chip \cite{Zheng2019}, would result in larger dispersive shifts and hence improved signal-to-noise ratios. 

\section{Data availability}
The data that support the findings of this study are available from the corresponding author upon reasonable request.

\section{Acknowledgements}
The work was supported by the European Union’s Horizon 2020 research and innovation program under Grant Agreements No. 688539 (MOS-QUITO), by the European Research Council (ERC) Projects No. 759388 (LONGSPIN) and and No. 810504 (QuCube), and by the French National Research Agency (ANR) projects MAQSi and CMOSQSPIN. S. Z. acknowledges support by an Early Postdoc.Mobility fellowship (P2BSP2\_184387) from the Swiss National Science Foundation.

\section{Author contributions}
R.E. and S.Z. performed the experiments with help from R.M.~. R.E., S.Z., R.M. and S.D.F. designed the experiment. L.H., B.B. and M.V. fabricated the sample. R.E. and S.Z. analyzed the results with inputs from V.M., J.L., A.A., Y.M.N., X.J., M.U., T.M., R.M. and S.D.F.~. V.M. and J.L. established the theoretical analysis and simulations under the supervision of Y.M.N.~. R.E., S.Z., V.M., R.M. and S.D.F. wrote the manuscript. M.S., X.J., M.V. and S.D.F. initiated the project.

\section{Competing Interests}
The authors declare no competing interests.

\section*{Appendices}
\appendix

\section{\label{apdx:theory}Theory}
\renewcommand\thefigure{A.\arabic{figure}} 
\setcounter{figure}{0} 

\subsection{Model of the driven double quantum dot}\label{Driven_dqd}

The double quantum dot system is resonantly driven by a fast electrical circuit with angular frequency $\omega_{\rm exc}$ and probed by an $LC$ circuit that is comparatively slow.
The dynamics of the double quantum dot is described by the Hamiltonian:
\begin{eqnarray}\label{H_dqd}
\nonumber H&=&\hbar\wRes(a^\dagger a+\frac{1}{2})\\
\nonumber&+&\sum_{i=1,2}\Big[\hbar g_{ci}n_i(a+a^\dagger)+\varepsilon_i(t)n_i+\frac{U_{i}}{2}n_{i}(n_{i}-1)\Big]\\
\nonumber&+&U_{m}n_{1}n_{2}-\sum_{\langle ij\rangle\sigma\sigma'}t_{i\sigma j\sigma'}c_{i\sigma}^\dagger c_{j\sigma'}\\
&+&\sum_{i=1,2}\frac{\mu_B}{2}(g_i {\bf B}_i)\cdot\sigma_i +H_{\kappa}+H_{\Gamma}.
\end{eqnarray}
Here $\wRes=1/\sqrt{L_{\rm{r}}C_{\rm{r}}}$ is the natural frequency of the probe $LC$ circuit, $a$ is the microwave photon operator, $\hbar g_{ci}=eV_{\rm{rms}}\alpha_{i}$ is the parameter of coupling between the dot charge and the quantum of the resonant circuit, with $\alpha_{i}$ the lever arm parameter between gate~1 and dot $i$, $V_{\rm{rms}}=\sqrt{\hbar\wRes/(2C_{\rm{r}})}$ the zero-point voltage of the $LC$ circuit \cite{Cottet2011}, and $e>0$ the elementary charge. $n_i=n_{i\downarrow}+n_{i\uparrow}$ is the occupation number of dot $i$, $U_{i}$ and $U_m$ are the intra- and inter-dot electrostatic energies,
 $\varepsilon_i(t)=e\alpha_{i}V_{G1}+e\beta_{i}(V_{G2}+V_{\rm exc}(t))$ is the time-dependent energy potential on dot $i$, with $V_{G1}$, $V_{G2}$ and $V_{\rm exc}(t)=V_{\rm exc}\cos(\omega_{\rm exc} t)$ the static and time-dependent voltages applied to gates~1 and 2 and $\beta_{i}$ the lever-arm parameter between gate~2 and dot $i$. $t_{i\sigma j\sigma'}$ are the spin-dependent charge tunneling matrix elements and $c_{i\sigma}$ are fermion operators for dot $i$ and pseudo-spin projection $\sigma$. $\mu_B$ is the Bohr magneton, $g_i$ and ${\bf B}_i$ are correspondingly the $g$-tensor and the magnetic field on dot $i$. $\sigma_i$ is the vector whose components are the operators $\sum_{\sigma\sigma'}c_{i\sigma}^\dagger\sigma_{\alpha\sigma\sigma'} c_{i\sigma'}$, with $\sigma_\alpha$ ($\alpha=x,y,\text{ and }z$) the Pauli matrices.
The terms $H_{\kappa}$ and $H_{\Gamma}$ represent the coupling to the environment of the $LC$ circuit and the double quantum dot.
We focus on the Pauli spin blockade configuration where the relevant charge states are of the type $\OneOne{}$ and $\TwoZero{}$.

\subsection{Two-state model}

Let us first analyze a two-state model that describes the dynamics of the double quantum dot in the vicinity of a single avoided crossing due to tunneling of a charge. 
We generalize to the multi-level double quantum dot in Section \ref{dqd_spectro}.
Tunneling couples the states with dot occupations $(M,N)$ and $(M+1,N-1)$ that are detuned by the electrostatic energy $\varepsilon=(\beta_1-\beta_2)eV_{G2}$.
Under driving, the time-dependent component of this detuning is $A_{\rm exc}\cos(\omega_{\rm exc} t)$, with $A_{\rm exc}=(\beta_1-\beta_2)eV_{\rm exc}$. 
Thus the two-state Hamiltonian writes
\begin{eqnarray}\label{H2l}
\nonumber H&=&-\frac{\varepsilon}{2}\sigma_{\rm{z}}-\frac{\Delta}{2}\sigma_{\rm{x}}-\frac{A_{\rm exc}}{2}\cos(\omega_{\rm exc} t)\sigma_{\rm{z}}\\
&+&\hbar\wRes (a^\dagger a+\frac{1}{2})+\hbar \gc\sigma_{\rm{z}}(a+a^\dagger)+H_{\kappa}+H_{\Gamma}.
\end{eqnarray}
Here we use the Pauli matrix representation in the space of the charge configurations so that $\sigma_{\rm{z}}\equiv n_1-n_2$, $\sigma_{\rm{x}}$ describes tunneling (which opens an energy gap $\Delta$), and the coupling parameter between the DQD charge and the microwave photon is 
\begin{equation}
 \hbar \gc=\frac{1}{2}(\alpha_{1}-\alpha_{2})e V_{\rm{rms}}.
\end{equation}
The Hamiltonian (\ref{H2l}) is formally equivalent to the theory of Ref.~\citenum{Hauss2008}. Our purpose is to adapt the model to the dispersive readout in our specific regime (Sec. \ref{phase_response}) and to apply it to the spectroscopy of the multi-level system (Sec. \ref{dqd_spectro}). 
In the rotating frame of the driven system Eq.~\ref{H2l} can be transformed \cite{Hauss2008} to:
\begin{eqnarray}\label{Ht2l}
\nonumber\tilde{H}&=&\frac{\hbar\wRabi}{2}\sigma_{\rm{z}}+\hbar\wRes (a^\dagger a+\frac{1}{2})+\hbar\tilde{\chi}\sigma_{\rm{z}}(a^\dagger a+\frac{1}{2})\\
&+&\hbar\tilde{g}(\sigma_+ a+\sigma_- a^\dagger)+H_{\kappa}+\tilde{H}_{\Gamma}.
\end{eqnarray}
Here the effective coupling parameter writes 
\begin{equation}
 \tilde{g}=\gc\cos\theta\frac{\wRabiZero}{\wRabi},
\end{equation}
and the off-resonance frequency shift is
\begin{equation}\label{chit}
 \tilde{\chi}=-\frac{2\hbar \gc^2\sin^2\theta}{E}\frac{\delta\omega}{\wRabi},\,\hbar\wRes\ll E.
\end{equation}
The above equations include the mixing factors $\cos\theta=\varepsilon/E$ and $\sin\theta=\Delta/E$, where $E=\sqrt{\varepsilon^2+\Delta^2}$ is the energy spacing of the levels near the avoided crossing. Furthermore $\wRabi=\sqrt{\delta\omega^2+\wRabiZero^2}$ is the full Rabi angular frequency, with $\delta\omega=\omega_{\rm exc}-E/\hbar$, $\wRabiZero=A_{\rm exc}\Delta/(2\hbar E)$, and $\tilde{H}_{\Gamma}$ represents the coupling of the system to the environment in the rotating frame.
We note that the off-resonance shift $\tilde{\chi}$ is taken in the adiabatic limit, which requires going beyond the rotating wave approximation \cite{Hauss2008,Zueco2009,Kohler2018,Park2020}. In our regime of parameters the interaction between the resonantly driven system and the probe circuit is well described by the rotating wave approximation (which is formally justified when $\omega_R-\wRes\ll\omega_R+\wRes$). The dot detuning range where this approximation is accurate is $\delta\varepsilon\approx A_{\rm exc}\Delta/(4\hbar\omega_r)>\Delta$. 

\subsection{Coupling to the environment}

As a model of dissipation we consider the coupling between the active charge in the double quantum dot and bosonic modes that can be phonons or other modes of the environment with regular noise spectra that couple to the charge. In the charge basis the coupling between the system and its environment is described by the Hamiltonian
\begin{equation}\label{HGamma}
 H_{\Gamma}=\sum_{\alpha}\hbar\omega_{\alpha}b_{\alpha}^\dagger b_{\alpha}+A\sigma_{\rm{z}},
\end{equation}
with $A=\sum_{\alpha}\lambda_{\alpha}(b_{\alpha}+b_{\alpha}^\dagger)$, $\alpha$ being the index of the mode of the environment.
Then the standard free evolution relaxation rates are
\begin{subequations}
\begin{align}
 &\Gamma_\downarrow=\frac{\sin^2\theta}{\hbar^2} S_{AA}(E/\hbar),\\
 &\Gamma_\uparrow=\frac{\sin^2\theta}{\hbar^2} S_{AA}(-E/\hbar),\\
 &\Gamma_\varphi=\frac{2\cos^2\theta}{\hbar^2} S_{AA}(0),
\end{align}
\end{subequations}
with $S_{AA}(\omega)$ the noise correlation function \cite{Clerk2010}.

\subsection{Master equations}

We describe the dynamics of the probed system semi-classically with the Bloch master equations:
\begin{subequations}
\begin{align}
 \nonumber&\langle\dot{\sigma}_-\rangle=-i\wRabi\langle\sigma_-\rangle-2i\tilde{\chi}\langle\sigma_-(a^\dagger a+1/2)\rangle\\
 &~~~~~~+i\tilde{g}\langle\sigma_{\rm{z}}a\rangle-\tilde{\Gamma}_2\langle\sigma_-\rangle,\label{dotsigma-}\\
 &\langle\dot{\sigma}_z\rangle=-2i\tilde{g}(\langle a\sigma_+\rangle-\langle a^\dagger\sigma_-\rangle)-\tilde{\Gamma}_1\langle\sigma_{\rm{z}}\rangle + \tilde{\Gamma}_\uparrow-\tilde{\Gamma}_\downarrow.\label{dotsigmaz}
\end{align}
\end{subequations}
The photon number dependent term in Eq.~\ref{dotsigma-} is negligible in the linear regime and is already included in Eq.~\ref{Ht2l} as an oscillator shift. Also $\tilde{\Gamma}_2=\tilde{\Gamma}_1/2+\tilde{\Gamma}_\varphi$ is the total dephasing rate of the dressed two-level system that includes the relaxation rate $\tilde{\Gamma}_1=\tilde{\Gamma}_\downarrow+\tilde{\Gamma}_\uparrow$ and the pure dephasing rate $\tilde{\Gamma}_\varphi$. The rate of decoherence of the driven system differs from the rate of decoherence in the absence of driving and it is found \cite{Ithier2005} to be
\begin{equation}
 \tilde{\Gamma}_2=\frac{3-\cos^2\eta}{4}\Gamma_1+\cos^2\eta\Gamma_\varphi+\frac{1}{2}\sin^2\eta\Gamma_\nu,
\end{equation}
where $\cos\eta=\delta\omega/\wRabi$ and $\sin\eta=\wRabiZero/\wRabi$, $\Gamma_1=\Gamma_\downarrow+\Gamma_\uparrow$ is the energy relaxation rate and $\Gamma_\nu$ is proportional to the spectral function of the noise at the Rabi frequency \cite{Ithier2005}.
Within the environment model taken here the rate $\Gamma_\nu$ evaluates to 
\begin{equation}
\Gamma_\nu=\frac{2\cos^2\theta}{\hbar^2} S_{AA}(\wRabi).
\end{equation}

For a relatively weak coupling $\gc$ the first term on the right-hand side of Eq.~\ref{dotsigmaz} can be neglected and the slow probe resonator does not significantly change the average population of the states. We note $D=-\langle\sigma_{\rm{z}}\rangle=\tilde{P}_--\tilde{P}_+$ the difference of the stationary occupation probabilities of the ground and excited states which becomes
\begin{equation}
 D=\frac{\tilde{\Gamma}_\downarrow-\tilde{\Gamma}_\uparrow}{\tilde{\Gamma}_1};\,4\gc^2 n_{ph}\ll\tilde{\Gamma}_1\tilde{\Gamma}_2,
\end{equation}
where $n_{ph}$ is the average number of photons in the $LC$ circuit.
In this regime, with the relaxation rates defined in \cite{Ithier2005}, we find
\begin{equation}\label{D}
 D=\frac{-\cos\eta(\Gamma_\downarrow-\Gamma_\uparrow)}{(1+\cos^2\eta)\Gamma_1/2+\sin^2\eta\Gamma_\nu}.
\end{equation}
This manifests population inversion in the dressed basis for $\cos\eta>0$ ($\delta\omega>0$) \cite{Hauss2008}. 

\subsection{Circuit phase response}\label{phase_response}

We use input-output theory \cite{GardinerZoller2004, Clerk2010, Schroer2012, Kohler2018} in order to calculate the phase shift of the signal due to its interaction with the DQD. In the input-output approach the dynamics of the circuit mode is given by the quantum Langevin equation \cite{GardinerZoller2004}. With Hamiltonian (\ref{Ht2l}) the equation reads in the rotating frame of the incoming microwave photon of angular frequency $\omega_p$ (which is close to the natural angular frequency of the resonant circuit $\wRes$):
\begin{equation}\label{HL}
\dot{a}=-i(\wRes-\omega_p)a-i\tilde{\chi}\sigma_{\rm{z}}a-i\tilde{g}\sigma_--\frac{\kappa}{2}a-\sqrt{\kappa}b_{\rm in},
\end{equation}
where $\kappa$ is the rate of escape of the photons and $b_{\rm in}$ is the incoming microwave photon field.

By solving Eq.~\ref{dotsigma-} at the frequency $\omega_p$ together with Eq.~\ref{D}, and using the input-output relations, we obtain the coefficient of reflection between the incoming and the outgoing microwave signals:
\begin{equation}\label{Ratio}
r(\omega_p)=1+\frac{i\kappa}{\wRes-\omega_p-i\frac{\kappa}{2}+\resp(\omega_p)},
\end{equation}
where $\chi(\omega_p)$ is the charge response function obtained at second order in the oscillator-DQD coupling:
\begin{equation}\label{Resp}
\resp(\omega_p)=-\Big(\tilde{\chi}+\frac{\tilde{g}^2}{\wRabi-\omega_p-i\tilde{\Gamma}_2}\Big)D.
\end{equation}

The real part of Eq.~\ref{Resp} yields the linear shift of the resonator frequency. When $\omega_p=\wRes$ and at linear order in the response function we get the phase shift 
$\delta\phi=4\operatorname{Re}\resp(\wRes)/\kappa=4Q\operatorname{Re}\resp(\wRes)/\wRes$, with $Q=\wRes/\kappa$ the quality factor of the $LC$ resonator.
 
\subsection{Double quantum dot spectroscopy}\label{dqd_spectro}

The above model naturally generalizes to multi-level systems, and we apply it to the spectroscopy of the double quantum dot.
We diagonalize the part of the Hamiltonian (\ref{H_dqd}) that does not include the drives and thus we obtain the energy levels and the corresponding states of the double quantum dot exactly. 
The mixing angles are: 
\begin{subequations}\label{angles}
\begin{align}
 &\sin\theta= \langle \gamma|(n_1-n_2)|\epsilon\rangle=2\langle \gamma|n_1|\epsilon\rangle=-2\langle\gamma|n_2|\epsilon\rangle,\\
 &\cos\theta=\langle \gamma|n_1|\gamma\rangle-\langle\epsilon|n_1|\epsilon\rangle=-\langle \gamma|n_2|\gamma\rangle+\langle\epsilon|n_2|\epsilon\rangle,
\end{align}
\end{subequations}
where $|\gamma\rangle$ is the ground state of the DQD with energy $E_\gamma$ and $|\epsilon\rangle$ an excited state of energy $E_\epsilon$.
To calculate the response of the multilevel system we sum Eq.~\ref{Resp} over all excited states.

\subsection{\label{apdx:multiphoton}Multi-photon processes}

The above analysis generalizes to multi-photon resonant processes. For the $n$-photon transition between the levels $\gamma$ and $\epsilon$ we substitute \cite{Ono2017}
\begin{eqnarray}
\delta\omega&\to&\delta\omega^{(n)}=n\omega_{\rm exc}-(E_\epsilon-E_\gamma)/\hbar,\\
\wRabiZero&\to&\wRabiZero^{(n)}=n\omega_{\rm exc}|\tan\theta|J_n\Big(\frac{A_{\rm exc}|\cos\theta|}{\hbar\omega_{\rm exc}}\Big),\\
\omega_{R}&\to&\omega_{R}^{(n)}=\sqrt{(\delta\omega^{(n)})^2+(\wRabiZero^{(n)})^2},
\end{eqnarray}
where $J_n$ is the n$^{\rm th}$ Bessel function of the first kind. For small argument it approximates as $J_n(z)\approx(z/2)^n/n!$ ($z\ll1$).

\begin{figure}[!htbp]
\centering
\includegraphics[width=\columnwidth]{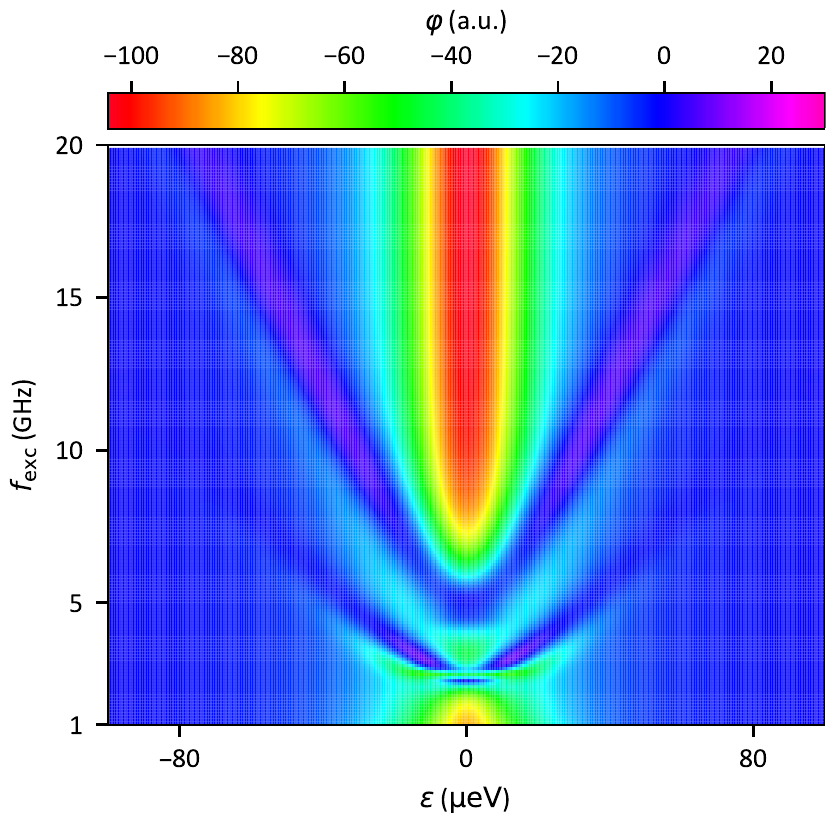}
    \caption{\label{fig:apdx:multiphotonth}\textbf{Phase response of the driven double quantum dot as given by Eqs. (\ref{Ratio}) and (\ref{Resp}) for the one and two photon processes at zero magnetic field.} The parameters are $Q=24$, $\hbar\wRes=\SI{2}{\micro\electronvolt}$, $\hbar g_c=\SI{0.15}{\micro\electronvolt}$. The double quantum dot is modeled by the charge tunneling $t=\SI{11.8}{\micro\electronvolt}$, tunneling with spin flip $t_{so}=\SI{8.3}{\micro\electronvolt}$, and g-factors $g_1=4$, $g_2=1.91$. Detuning energy $\varepsilon$ is counted from the point of charge degeneracy at zero magnetic field. The amplitude of the resonant drive is $A_{\rm exc}=\SI{15}{\micro\electronvolt}$. The phonon equilibrium temperature is significantly smaller than all the energy separations between the levels, which corresponds to the limit $\Gamma_\uparrow\ll\Gamma_\downarrow$. With our environment model we set for simplicity $S_{AA}(E)/\hbar=\SI{0.5}{\micro\electronvolt}$ and $S_{AA}(0)/\hbar=S_{AA}(\wRabi)/\hbar=\SI{0.5}{\micro\electronvolt}$. To account for quasistatic noise due to the electrical environment we furthermore convolute the computed signal with a Gaussian function of width $\sigma=\SI{5}{\micro\electronvolt}$.}
\end{figure}

The total response is obtained by summing over all resonant photon processes $n=1,2,\dots$. In Fig.~\ref{fig:apdx:multiphotonth} we show the calculated response up to $n=2$.

\section{\label{apdx:extractions}Extractions}

\subsection{\label{apdx:tunnel}Tunnel coupling \& spin-orbit interaction}
In order to get a more precise and correct extraction of the tunnel coupling and the spin-orbit interaction, we need to take into account the charge noise fluctuations impact into our fitting model.\\

The transition frequency ($f$) of a charge qubit as a function of detuning ($\varepsilon$) is:
\begin{equation}
    \label{eq:transition_frequency}
    f = \frac{\sqrt{\varepsilon^2 + \Delta^2}}{h},
\end{equation}
where $h$ is the Planck constant and $t$ is the tunnel coupling giving rise to an anticrossing gap $\Delta$.

Low frequency charge noise will lead to fluctuations of the qubit transition frequency. Assuming that the noise on the detunig axis is Gaussian, its probability density function can be written as:
\begin{equation}
    \label{eq:pdf_epsilon}
    \rho_\varepsilon(\varepsilon) = \frac{1}{\sigma_\varepsilon \sqrt{2\pi}} \cdot e^{-\frac{1}{2}(\frac{\varepsilon}{\sigma_\varepsilon})^2},
\end{equation}
where $\sigma_\varepsilon$ is the standard deviation of the distribution of detuning noise. To transform the probability density function of the detuning noise into the probability density function for the transition frequency, we use the following relation:
\begin{equation}
    \label{eq:pdf_transformation}
    \rho_y(y) = \rho_x(x(y))\cdot \dfrac{dx(y)}{dy},
\end{equation}
where $\rho_y(y)$ is the probability density function of $y$ and $y(x)$ is a function of $x$, whose probability density function is given by $\rho_x(x)$.\\

Therefore, we find the following probability density function for the transition frequency:
\begin{equation}
\rho_f(f) =
\begin{cases} 
    0 &, f\leq \Delta/h \\ 
    \frac{2h^2f}{\sigma_\varepsilon\sqrt{2\pi}}\cdot \frac{e^{-\frac{1}{2}(\frac{(hf)^2 - \Delta^2}{\sigma_\varepsilon^2})}}{\sqrt{(hf)^2 - \Delta^2}} &, f>\Delta/h
\end{cases}
\end{equation} 

The factor of 2 in $\rho_f(f)$ comes from the fact that there are two solutions to Eq.~\ref{eq:transition_frequency} for $\varepsilon(f)$.

The linewidth of the qubit transition is given by the lifetime of the excited state (assuming that the ground state cannot decay and has infinite lifetime). From Heisenberg uncertainty principle, we know that $\Delta E\Delta t \geq \hbar$ \cite{demtroder2016experimentalphysik}. This directly translates into a spectral width $\Delta f = 1/(2\pi T_1)$, where $T_1$ is the lifetime of the excited state. Then, the full linewidth is described by a Lorentzian centered around frequency $f_0$:
\begin{equation}
    \label{eq:Lorentzian}
    \mathcal{L}(f) = A \cdot \frac{(\frac{\Gamma}{2})^2}{(f - f_0)^2 + (\frac{\Gamma}{2})^2},
\end{equation}
where A is the amplitude of the spectral ray at $f=f_0$ and $\Gamma=\frac{1}{T_1}$ is the full width at half maximum and the lifetime of the excited state.

The final lineshape of the phase response as a function of drive frequency at an anticrossing is given by the convolution of the intrinsic lineshape of the qubit (Lorentztian characterized by $T_1$) and the probability density function of the transition frequencies \cite{Penfold-Fitch2017}:
\begin{equation}
    \label{eq:convolution}
    \varphi(f) = \int_{-\infty}^{+\infty} \mathcal{L}(\nu)\cdot \rho_f(f-\nu)\cdot d\nu + \varphi_0,
\end{equation}
where $\varphi_0$ is a phase offset.

We compute Eq.~\ref{eq:convolution} numerically and pass it as the fitting model in order to extract the tunnel coupling at the anticrossing.

\begin{figure}[!htbp]
\centering
\includegraphics[width=\columnwidth]{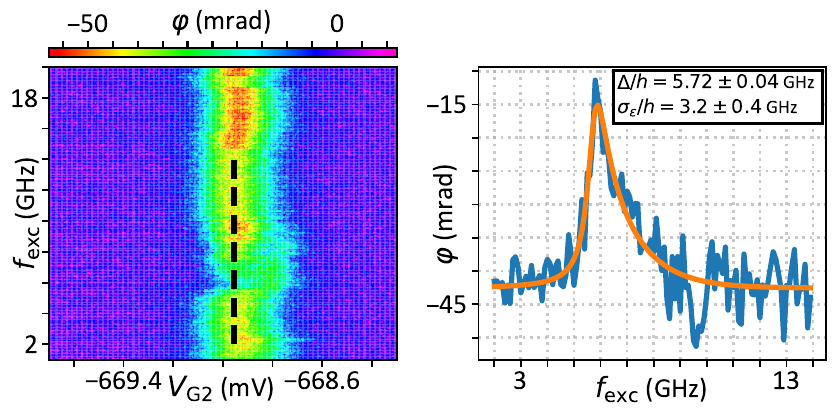}
    \caption{\label{fig:apdx:tcextraction}\textbf{Tunnel coupling extraction for ICT~2.} Phase response of the LC resonator as a function of $\vgTwo{}$ and $\fExc{}$ is shown on the left and a line-cut (black dashed line) is shown on the right. The line-cut is fitted with Eq.~\ref{eq:convolution} to extract the framed parameters.
    }
\end{figure}

\begin{figure}[!htbp]
\centering
\includegraphics[width=\columnwidth]{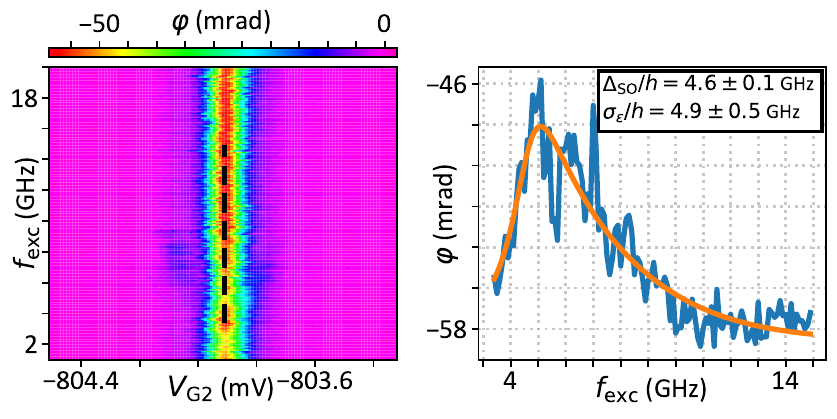}
    \caption{\label{fig:apdx:tsoextraction}\textbf{Spin orbit interaction extraction for ICT~1.} Phase response of the LC resonator as a function of $\vgTwo{}$ and $\fExc{}$ is shown on the left and a line-cut (black dashed line) is shown on the right. The line-cut is fitted with Eq.~\ref{eq:convolution} to extract the framed parameters.
    }
\end{figure}

The fittings that allowed the extraction of $\Delta$ (in the case of ICT~2) and $\deltaSO{}$ (in the case of ICT~1) are presented in Fig.~\ref{fig:apdx:tcextraction} and Fig.~\ref{fig:apdx:tsoextraction}, respectively. The dashed line cuts along the anticrossing detuning were taken from spectroscopy maps measured at the lowest constant room temperature power.

\subsection{\label{apdx:alpha}Alpha factor}

By following the center of the dip-peak structure of the wings, we get the energy gap value between the ground state and the excited state for each detuning value.

\begin{figure}[!htbp]
\centering
\includegraphics[width=\columnwidth]{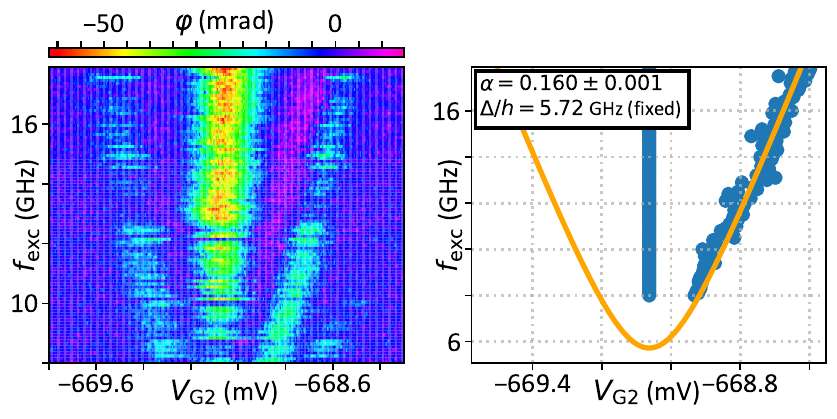}
    \caption{\label{fig:apdx:alphaextraction}\textbf{Alpha factor extraction for ICT~2.} Phase response of the LC resonator as a function of $\vgTwo{}$ and $\fExc{}$ (with power calibration) is shown on the left. The scatter plot on the right retraces the positions of the central line as ell as the right dip-peak structure from the left graph. The orange curve is a fit using Eq.~\ref{eq:transitionfrequency} which yields the framed results.
    }
\end{figure}

Fig.~\ref{fig:apdx:alphaextraction} illustrates the extraction of the $\alpha$-factor. Starting from the calibrated spectroscopy map at zero magnetic field, we mark both the position of the central dip and the center of wing dip-peak. We align afterwards the positions corresponding to zero detuning around their mean value $V_0$. We can then fit alpha to the obtained wing positions with:
\begin{equation}
    \label{eq:transitionfrequency}
    h\fExc{} = \sqrt{\alpha^2(\vgTwo{}-V_0)^2 + \Delta^2},
\end{equation}
where $\Delta$ is the anticrossing gap extracted as in section \ref{apdx:tunnel} and input in the model as a fixed parameter.

\subsection{\label{apdx:gcoupling}Charge photon coupling}

\begin{figure}[!htbp]
\centering
\includegraphics[width=\columnwidth]{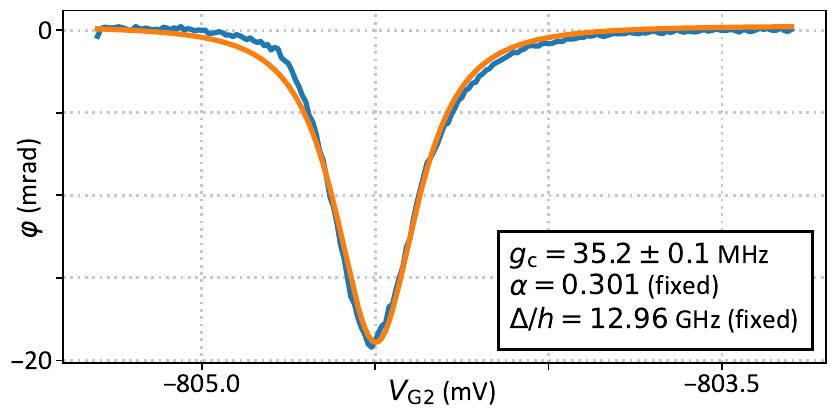}
    \caption{\label{fig:apdx:gcextractionone}\textbf{Charge photon coupling extraction for ICT~1.} The experimental data (blue) is fitted (orange) with Eq.~\ref{eq:phase} and Eq.~\ref{eq:dispersive_shift_two_state} to extract the relevant parameters. 
    }
\end{figure}

\begin{figure}[!htbp]
\centering
\includegraphics[width=\columnwidth]{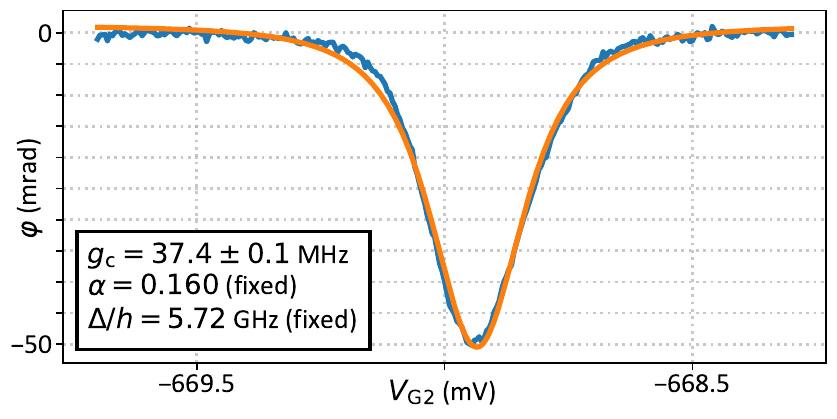}
    \caption{\label{fig:apdx:gcextractiontwo}\textbf{Charge photon coupling extraction for ICT~2.} The experimental data (blue) is fitted (orange) with Eq.~\ref{eq:phase} and Eq.~\ref{eq:dispersive_shift_two_state} to extract the relevant parameters. 
    }
\end{figure}

The charge photon coupling is extracted using Eq.~\ref{eq:phase} and Eq.~\ref{eq:dispersive_shift_two_state} as a model. Each experimental dip in Fig.~\ref{fig:apdx:gcextractionone} and \ref{fig:apdx:gcextractiontwo} is a 40 time average of the same measurement at zero magnetic field with no spectroscopy drive. We then fit these data while introducing the already extracted entities ($\alpha$, $t$, $Q$ \dots) as fixed parameters.\\
The decoherence rate $\Gamma$ is neglected here since we assume that it is small compared to the tunnelling.

\subsection{\label{apdx:gfactor}$g$-factors}

Due to $g$-factor difference between the two quantum dots, the Zeeman splitting induced by the applied magnetic field is not the same for both spins. 

\begin{figure}[!htbp]
\centering
\includegraphics[width=\columnwidth]{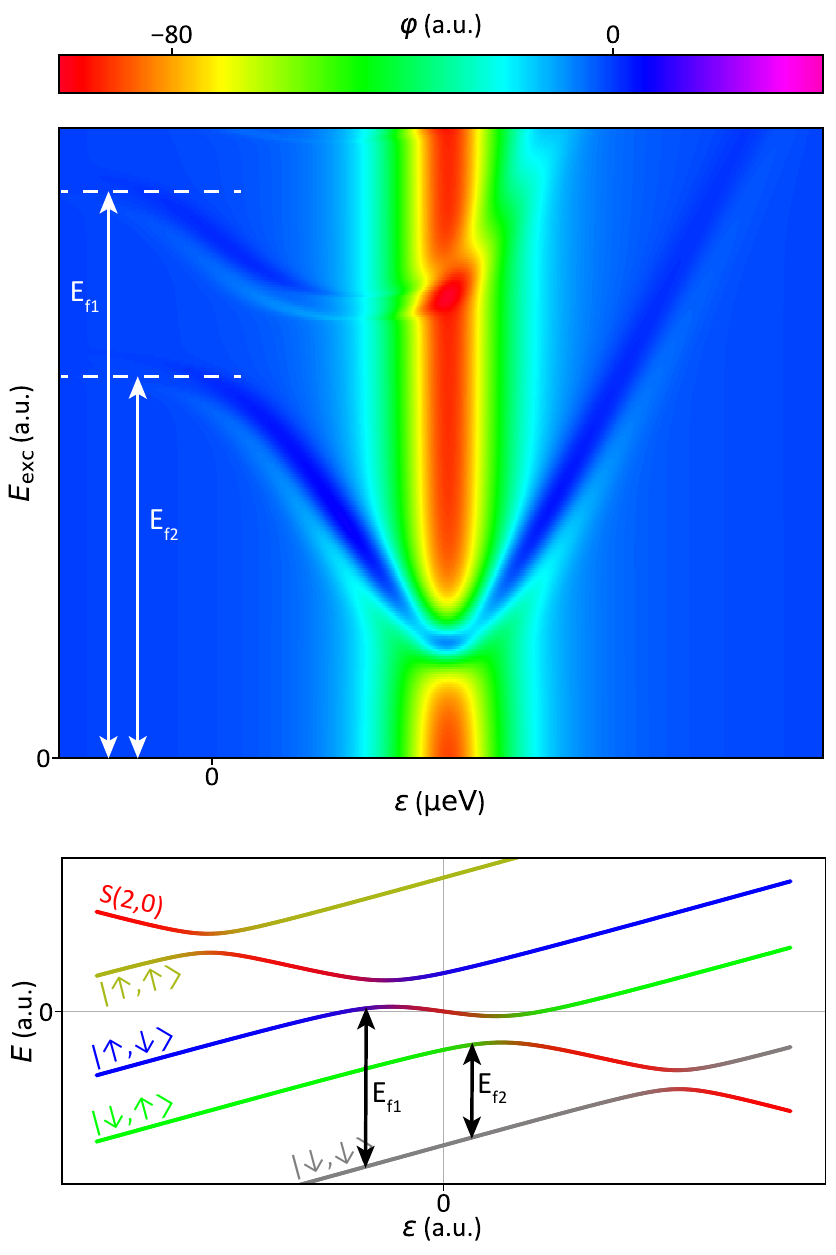}
    \caption{\label{fig:apdx:gfactors}\textbf{$g$-factors extraction.} Simulated response of the driven DQD system at finite magnetic field in the case of an even charge parity (top panel) and the corresponding schematic of the energy diagram (bottom panel) highlighting the transition energies $E_{\rm{f1}}$ and $E_{\rm{f2}}$ necessary for $g$-factors extraction.
    }
\end{figure}

The energies $E_{\rm{f1}}$ and $E_{\rm{f2}}$ illustrated in Fig.~\ref{fig:apdx:gfactors} represent the energies necessary to flip respectively the spin of QD 1 and the spin of QD 2. These energies give access to the $g$-factors knowing the applied magnetic field $B$ since:
\begin{align}
    \label{eq:gfactors}
        E_{\rm{f1}} &= g_1 \mu_B B\\
        E_{\rm{f2}} &= g_2 \mu_B B
\end{align}

$E_{\rm{f1}}$ and $E_{\rm{f2}}$ are marked on a theoretical spectroscopy map on the top panel of Fig.~\ref{fig:apdx:gfactors}.

\bibliography{main.bib}

\end{document}